\documentclass{PoS}

\title{Studying Supernovae in the Near-Ultraviolet with the \\
NASA {\it Swift} UVOT Instrument}

\ShortTitle{Swift UV}

\author{\speaker{Peter A. Milne}\\
        University of Arizona\\
        E-mail: \email{pmilne@as.arizona.edu}}
\author{P.J. Brown\\
        University of Utah\\
        E-mail: \email{pbrown@physics.utah.edu}}

\abstract{Observations in the near- and mid-ultraviolet (NUV: 2000--3500$\AA$)
 performed with the NASA {\it Swift} UVOT 
instrument have revealed that optically-normal SNe Ia feature NUV-optical 
color evolution that can be divided into NUV-blue and NUV-red groups, 
with roughly one-third of the observed events exhibiting NUV-blue color 
curves. Combined with an apparent correlation between NUV-blue events and the 
detection of unburned carbon in the optical spectra, the grouping might 
point to a fundamental difference within the normal SN Ia classfication. 
Recognizing the dramatic temporal evolution of the NUV-optical colors for 
all SNe Ia, as well as the existence of this sub-division, is important for 
studies that compare nearby SNe Ia with intermediate or high-$z$ events, for 
the purpose of the cosmological utilization of SNe Ia. SN 2011fe is shown to 
be of the NUV-blue groups, which will be useful towards interpretation 
of the gamma-ray line results from the INTEGRAL SPI campaign on SN 2011fe.}

\FullConference{The Extreme and Variable High Energy Sky - extremesky2011,\\
		September 19-23, 2011\\
		Chia Laguna (Cagliari), Italy}

\begin{document}

\section{UVOT Color Curves of Normal SNe Ia}

SNe Ia are luminous events that synthesize an
appreciable fraction of the iron-peak elements in the universe 
(Iwamoto et al. 1999)
and can be utilized to measure large distances. SN Ia distance estimates 
have revealed that the expansion of the universe is 
accelerating (Riess et al. 1998; Perlmutter et al. 1999).
There is considerable interest in extending the current utilization of 
SNe Ia, both by better understanding the SN Ia event and by widening the
rest-frame wavelength ranges being utilized.
The near-ultraviolet wavelength (NUV: 2000--3500$\AA$) range, 
including the range covered
by the U-band, has posed challenges for the efforts to understand
type Ia supernovae (SNe Ia).

The UVOT instrument on the NASA {\it Swift} mission possesses
the ability to quickly schedule observations of a new supernova,
increasing the chances of observing the important early epochs.
The 30cm UVOT telescope employs 3 NUV filters ($uvw1$, $uvm2$, $uvw2$) and
three optical filters ($u$,$b$,$v$), and SNe are typically observed with
all six filters at all epochs. Brown et al. (2009) presented photometry
from the first 2.5 years of UVOT observations, 2243 total observations
of 25 SNe. The photometry presented in
Brown et al. (2009) was studied
for light curve properties in Milne et al. 2010 (hereafter M10) and for absolute
magnitudes in Brown et al. 2010 (hereafter B10). Among the findings reported in 
those works, it was shown that the color evolution of the UVOT-$u$ and
NUV filters relative to the optical ($V$) band was both dramatic and
relatively homogeneous within the sub-class of normal 
SNe Ia.\footnote{We employ the definition of normal SNe~Ia to be events with 
broad light curves peaks, and thus not ``narrow-peaked" and being 
photometrically and spectroscopically normal, and thus not 
02cx-like, 00cx-like, 08ha-like, etc..} 
The colors become rapidly bluer until $\sim$6 days pre-peak in the
$B$ filter, then abruptly become redder from $\sim$20 days. There
was a lone significant outlier to the color evolution, SN 2008Q, 
which featured a similar shape to the color evolution, offset to 
bluer NUV-optical colors at all early epochs. Screening 2008Q, 
the low-scatter suggested that the rest-frame $U$ band might be a 
useful filter for standard
candle applications, since the rest-frame $V$ filter is a primary filter in
standard candle studies. The absolute magnitudes of the peak
magnitudes of the UVOT-$u$ light curves were able to be fit with a
linear luminosity-width relation (LWR). The scatter about that $U$ band
LWR was similar to what was seen in the $B$ and $V$ filters. The
LWR for the $uvw1_{rc}$ filter had somewhat larger scatter and the
$uvw2_{rc}$ and $uvm2$ filters had large scatter, suggesting that 
the cosmological utilization of SNe~Ia might only extend, at farthest, 
to the wavelength range sampled by the $uvw1$ filter.\footnote{$uvw1_{rc}$  and 
$uvw2_{rc}$ refer to ``red-tail corrected" $uvw1$ and $uvw2$ filters. 
See B10 for details of the method used to try to account for 
contamination of those filters from longer wavelength emission.} 

New SNe Ia have been observed since M10 and B10. In particular, more
events were observed that featured colors similar to those seen in
SN 2008Q. The colors curves for 21 normal SNe~Ia for which template-subtracted 
photometry has been obtained, is shown in Figures 1 \& 2, where Figure 2 is a 
zoomed in version of Figure 1. The basic features 
shown in M10 are present for this larger sample, with the added recognition 
that 6 of 21 SNe are bluer in the NUV-$v$ colors. The NUV-blue events 
appear to follow an evolution that is similar to the larger collection 
of NUV-red events, but for being offset by $\sim$0.4 magnitudes. 

\section{NUV Colors versus Unburned Carbon}

The recognition that roughly one-third of the optically normal SNe Ia were 
appreciably bluer in the NUV-$v$ colors led to the obvious 
search for correlation with parameters derived from the optical 
observations such as the light curve peak-width (i.e. $\Delta m_{15}(B)$ Phillips 1993), 
the blueshift of spectral features (i.e. HV/N, Wang et al. 2009), or 
time-variation of the blueshift of spectral features (i.e. HVG/LVG, Benetti et al. 2005),
The Nearby Supernova Factory team, concentrating on the search for 
unburned carbon in the optical spectra, first recognized that all 
the UVOT SNe Ia that feature blue NUV-optical colors also had featured 
unburned carbon in the early-epoch spectra with SNe Ia (Thomas et al. 2011). 
This potential correlation presents an interesting opportunity to 
directly tie NUV emission to explosion physics, as detectable unburned carbon 
is possibly the signature of a burning front that fails farther from the 
surface than normal. The optical signature of unburned carbon becomes 
difficult to detect with epoch, typically disappearing before optical peak. 
If the NUV-blue/carbon correlation holds up, the NUV-optical colors might 
prove the best way to distinguish this variety of explosion.

Two other papers have been published that explore the search for unburned 
carbon features in optical spectra, the Parrent et al. (2011) analysis of 
archival SN spectra, and the Folatelli et al. (2011) analysis of Carnegie 
Supernova Project ($CSP$) spectra. The two papers increase the sample of 
events for which UVOT colors can be compared with the carbon search. Further, 
the $CSP$ paper quantified the time-evolution of the pseudo equivalent width of 
the carbon feature, which can eventually be used to improve from a binary, yes/no 
determination of the existence of carbon in future UVOT works (Milne et al., in 
preparation). 

\section{SN 2011fe and Iron-peak Nucleosynthesis}

{\it Swift} UVOT observed SN~2011fe extensively, with observations being 
initiated by the Palomar Transient Factory upon recognition that there was a 
new supernova in M101 (Nugent et al. 2011). 
The UVOT observations provided solid constraints 
upon the explosion epoch of the SN, but perhaps more importantly upon  
early-epoch emission from the interaction of the SN shock with potential companion 
stars. That study concluded that a Main Sequence companion at a distance of 
a few solar radii experiencing Roche Lobe overflow was excluded at a 95\% confidence 
level for realistic viewing angles 
(Brown et al. 2012). The UVOT campaign continues into 2012, allowing SN~2011fe to 
be compared with the other normal SNe~Ia. Figures 1 \& 2 show the color curves 
of SN~2011fe plotted with other normal, low-extinction SNe Ia.
The SN~2011fe photometry employed special observing modes and reduction 
to account for coincidence loss, but the SN observations were still saturated near peak in 
the optical filters. This accounts for the gap in color curves in Figures 1 and 2. 
SN~2011fe is clearly of the NUV-blue category, and unburned carbon has been detected
in the optical spectra (Nugent et al. 2011). This is important to mention as
the INTEGRAL campaign that was executed on SN~2011fe may provide the best
detection of gamma-ray line emission in a SN Ia due to $^{56}$Ni decay 
products.\footnote{Preliminary analysis suggests only upper limits, 
both for $^{56}$Ni gamma-ray lines, Isern et al. 2011a, 
and for $^{56}$Co gamma-ray lines, Isern et al. 2011b.}  
Wang et al. (2011) presented NUV and optical photometry and early-epoch spectra of
four SNe Ia, utilizing HST and ground-based telescopes.
One SN in that sample, SN~2004dt, was NUV-blue, permitting a spectroscopic
study of a NUV-blue event. The spectra reveal an excess in the 2900--3500$\AA$
wavelength range, when compared with spectra of the other SN Ia events. Emission 
in that wavelength range is typically atributed to the degree of absorption
from iron-peak elements, so the NUV excess for the NUV-blue events might point
to relatively little iron-peak absorption. If this is caused by a low abundance
of iron-peak elements near the surface of the ejecta SN~2011fe might feature weaker 
gamma-ray line emission than might be seen in NUV-red events. It is important to point 
out that there are many alternative explanations for the NUV excess (see Wang et al. (2011)).  
This possibility will be interesting to explore as theoretical 
models attempt to explain the multi-wavelength evolution of the emission from 
this historic SN Ia event.

\begin{figure}
\includegraphics[width=.95\textwidth]{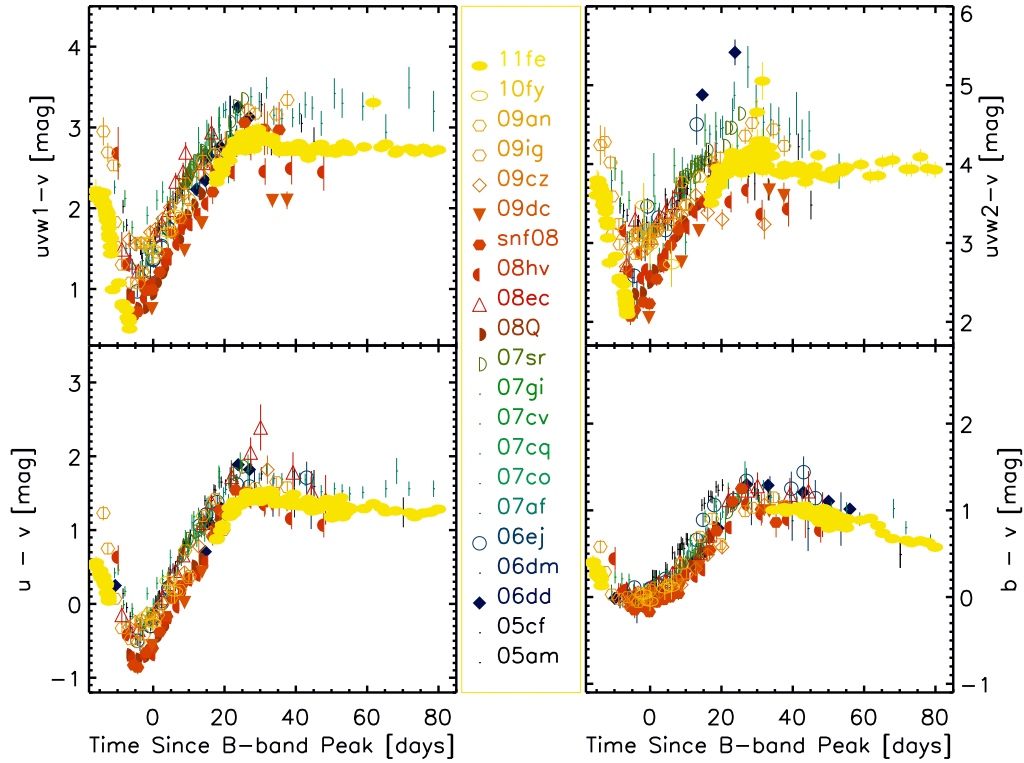}
\caption{Colors of 21 normal SNe~Ia compared to the
$v$ band. SNe Ia featured in M10 are shown with dots, SNe Ia new to 
this paper are plotted as labeled. The slope of the NUV-$v$ color changes 
are steeper, cover a larger total change and switch to a reddening trend 
more abruptly than the $b-v$ colors. The ``NUV-blue" group is shown
with filled symbols, the ``NUV-red" group with open symbols.}
\label{col_N}
\end{figure}

\begin{figure}
\includegraphics[width=1.0\textwidth]{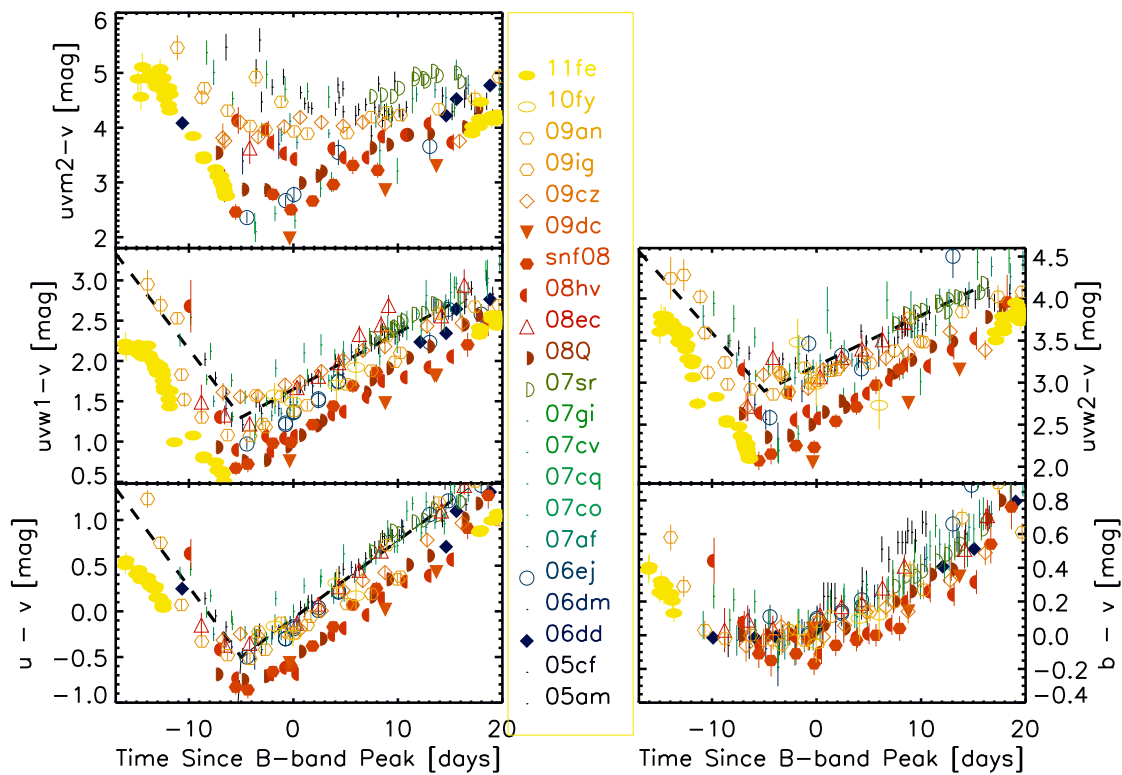}
\caption{Near-peak colors of 21 normal SNe~Ia compared to the
$v$ band. Plotting symbols the same as Figure 1, with the addition 
of $uvm2-v$ color curves. Pairs of dashed lines forming a ``V" are shown 
for visual comparison for $u-v$,$uvw1-v$,$uvw2-v$.}
\label{col_N_zoom}
\end{figure}

\end{document}